\documentclass[conference,manuscript]{IEEEtran}
\IEEEoverridecommandlockouts

\usepackage[justification=centering]{caption}
\usepackage{caption}
\usepackage{CJKutf8}
\usepackage{multicol}
\usepackage{bm}
\usepackage{amsmath,amssymb,amsfonts}
\usepackage{makecell}
\usepackage{CJKutf8}
\usepackage{tcolorbox}
\tcbuselibrary{breakable}
\usepackage[font=tiny]{subfig}
\usepackage{graphicx}
\usepackage{caption}
\usepackage{overpic}
\usepackage{balance}
\usepackage{booktabs}

\newcommand{\model}{\textsc{GeneSum}}

\def\BibTeX{{\rm B\kern-.05em{\sc i\kern-.025em b}\kern-.08em
    T\kern-.1667em\lower.7ex\hbox{E}\kern-.125emX}}
\begin{document}

\title{\model: Large Language Model-based Gene  Summary Extraction}

\author{\IEEEauthorblockN{Zhijian Chen$^{1,2,\S}$\thanks{$^\S$ Equal Contributions.}, Chuan Hu$^{1,2,\S}$, Min Wu$^4$, Qingqing Long$^{1,2}$, Xuezhi Wang$^{1,2}$, \\Yuanchun Zhou$^{1,2,3}$, Meng Xiao$^{1,*}$\thanks{$^*$ Corresponding author: shaow@cnic.cn.}}
\IEEEauthorblockA{\textit{$^1$Computer Network Information Center, Chinese Academy of Sciences, China} \\
\textit{$^2$University of Chinese Academy of Sciences, China} \\
\textit{$^3$Hangzhou Institute for Advanced Study, University of Chinese Academy of Sciences, China}\\
\textit{$^4$Institute for Infocomm Research (I$^2$R), Agency for Science, Technology and Research (A*STAR), Singapore}}}

\maketitle
\begin{abstract}
Emerging topics in biomedical research are continuously expanding, providing a wealth of information about genes and their function. 
This rapid proliferation of knowledge presents unprecedented opportunities for scientific discovery and formidable challenges for researchers striving to keep abreast of the latest advancements.
One significant challenge is navigating the vast corpus of literature to extract vital gene-related information, a time-consuming and cumbersome task.
To enhance the efficiency of this process, it is crucial to address several key challenges: (1) the overwhelming volume of literature, (2) the complexity of gene functions, and (3) the automated integration and generation.
In response, we propose \model, a two-stage automated gene summary extractor utilizing a large language model (LLM).
Our approach retrieves and eliminates redundancy of target gene literature and then fine-tunes the LLM to refine and streamline the summarization process. 
We conducted extensive experiments to validate the efficacy of our proposed framework.
The results demonstrate that LLM significantly enhances the integration of gene-specific information, allowing more efficient decision-making in ongoing research.

\end{abstract}

\begin{IEEEkeywords}
Gene summary, prompt learning, large language model. 
\end{IEEEkeywords}

\maketitle

\section{Introduction}
In recent years, genomic research has documented extensive data on gene functions, characteristics, and expressions in various species, significantly advancing our understanding. 
However, the extraction and summary of specific gene knowledge from this burgeoning literature remains a daunting, labor-intensive task, and is mainly carried out by experts. 
For example, the Entrez Gene database at the National Center for Biotechnology Information (NCBI)~\cite{federhen2012ncbi}, which stores comprehensive gene summary information, required numerous researchers to develop and maintain. 
In addition, the vast majority of genes lack succinct and descriptive summaries~\cite{schoch2020ncbi}.
Implementing automation in gene summarization can simplify this procedure, complement the knowledge base, and enable biologists to quickly comprehend essential information about target genes. 
Existing literature has partially address the gene summary problem.
(1) \textit{extractive-summarization}, which has been the traditional approach due to its straightforward methodology of selecting key sentences directly from texts~\cite{shang2014learning,chen2018fast}.  
However, it often results in summaries that are somewhat disjointed and may lack overall coherence if the extracted sentences do not flow naturally together.
(2) \textit{generative-summarization}, addresses these limitations by synthesizing new content that is not only faithful to the original information but also more cohesive and concise~\cite{gehrmann2018bottom}. 
However, the performance of these generative models is often constrained by the capabilities of the underlying neural networks, which can impact the accuracy and depth of the generated summaries.

To enhance the efficiency of this process, addressing the following challenges is essential:
\textbf{(C1)} Overwhelming Volume of Literature: The sheer quantity of publications makes it difficult to identify and assimilate key knowledge~\cite{xiao2021expert,xiao2023hierarchical, xiao2023interdisciplinary} about specific genes. 
\textbf{(C2)} Complexity of Gene Functions: Genes often have multiple functions and are involved in various pathways. 
\textbf{(C3)} Integration of Gene Functions with Literature Knowledge: Effectively combining detailed gene function descriptions with insights derived from the literature is essential for forming a complete picture of gene roles and interactions. 

\begin{figure*}[!ht]
\centering
\includegraphics[width=0.89\linewidth]{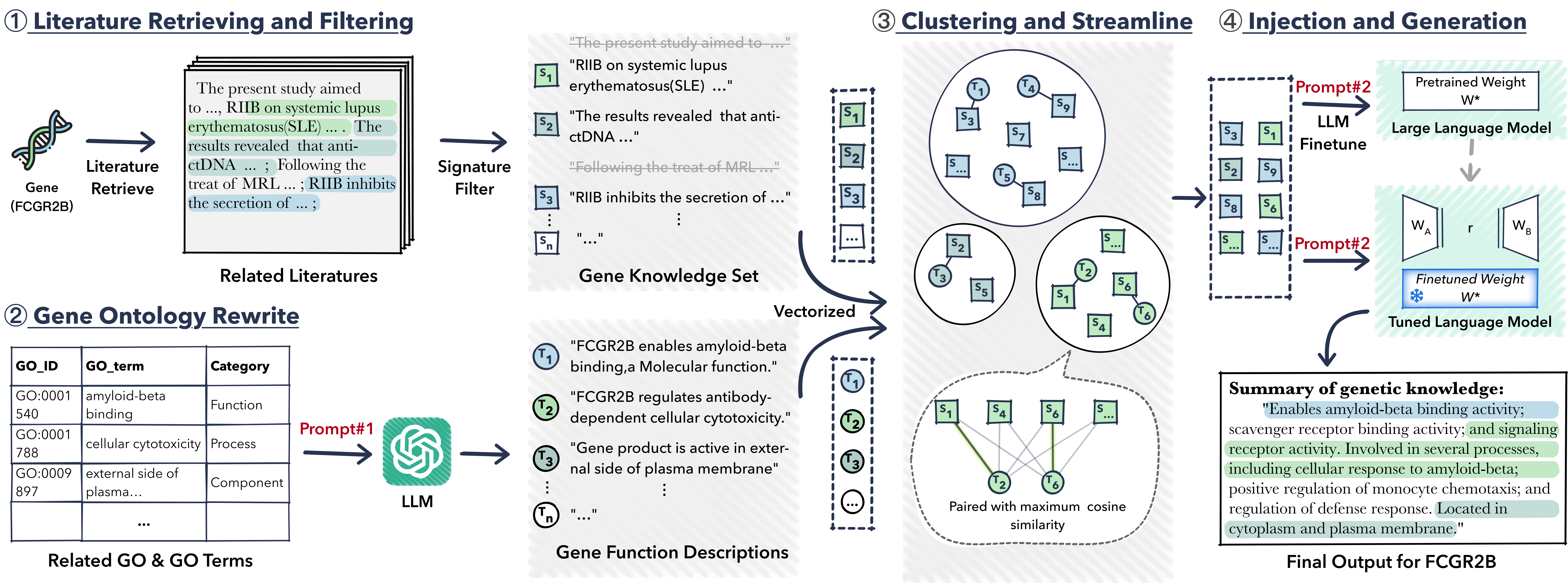}
\caption[width=\linewidth]{An overview of our framework.}
\label{model_overview}
\vspace{-0.5cm}
\end{figure*}

\textbf{Summary of Technical Contributions}: To achieve this, we introduce \textbf{Gene} \textbf{Sum}mary Extractor (\model), a two-stage automated gene summary extractor powered by a large language model (LLM). 
Initially, our system retrieves literature relevant to the target gene and analyzes the inherent relationships among knowledge entities to eliminate redundant content. 
This step ensures that only the most relevant and unique information is processed. 
We then fine-tune the LLM to enhance and streamline the summarization process, producing concise and informative summaries that effectively synthesize gene functions and literature insights. 
The contributions can be listed as follows:
\begin{itemize}
    \item \textbf{Innovative Formulation and Application}: We have defined the gene summary problem within the context of modern bioinformatics and are the first to apply LLMs to this challenge.
    \item \textbf{Advancement in Data Handling Techniques}: We have developed sophisticated data preprocessing techniques that significantly enhance the efficiency and accuracy of information extraction from genetic databases. 
    \item \textbf{Empirical Validation}: Through comprehensive experiments and case studies on real-world datasets, we demonstrate the effectiveness of our framework. 
\end{itemize}




\section{LLM-based Gene Summary Generation Framework}


\subsection{Literature Retrieving and Filtering.}

\noindent\textbf{The Importance of Signature Filtering.} 
Various kinds of texts often possess unique signature terms~\cite{lin2000automated,jin2009towards}. 
For instance, the frequent occurrence of words like \textit{sweat}, \textit{competition}, and \textit{racing} in a corpus may suggest that the topic is related to sports or competitions. 
Likewise, gene knowledge summaries have their own signature terms, particularly those that appear more often. 
These terms can be identified by comparing their expected versus observed frequencies.

\noindent\textbf{Signature Terms Filtering Method.}
We use the Pearson’s chi-square test~\cite{manning1999foundations} to extract topic signature terms from reference summaries in the training set by comparing the occurrence of terms in reference summaries with that of the randomly selected biological literature. 
Let $R$ denote the set of reference summaries in the training set and $\tilde{R}$ denote the set of randomly selected Biological literatures. The \textit{null hypothesis} and \textit{alternative hypothesis} are as follows: \\
\begin{equation}
\begin{aligned}
    & \mathrm{H}_{0}: P(t_{i} \mid R) = p = P(t_{i} \mid \tilde{R}) \\
    & \mathrm{H}_{1}: P(t_{i} \mid R) = p_{1} \neq p_{2} = P(t_{i} \mid \tilde{R})
\end{aligned}
\end{equation}

Under the \textit{null hypothesis}, the item $t_{i}$ appears with equal probability in both $R$ and $\tilde{R}$ . and  $t_{i}$ is independent of $R$. 
In contrast, the alternative hypothesis indicates that the term $t_{i}$ is correlated with $R$.
In this context, $H_{0}$ states that term $t_{i}$ is not a signature term, while $H_{1}$ proposes that term $t_{i}$ is a high-frequency signature term.

We then construct the following 2-by-2 contingency table.
 
The Pearson’s chi-square statistic is computed by\\
\begin{equation}
\begin{aligned}
 X^{2}=\sum_{i, j=1}^{2} \frac{\left(O_{i j}-E_{i j}\right)^{2}}{E_{i j}}
\end{aligned}
\end{equation}
where  $O_{i j}$  is the observed frequency and  $E_{i j}$  is the expected frequency.
According to the chi-square calculation formula, 
a larger  $X^{2}$ value indicates that the observed frequency is far from the expected frequency, suggesting rejection of the null hypothesis that $t_{i}$ is a signature term. 
A smaller $X^{2}$ value indicates that the observed frequency is close to the expected frequency, supporting the hypothesis that  $t_{i}$ is not a signature term.
We retain signature terms with larger $X^{2}$ to form a set of signature terms, and sentences containing less than three signature terms will be filtered~\cite{jin2009towards}. 
Finally, a set of candidate sentences is formed.

\begin{figure}[!h]
\centering
\includegraphics[width=0.9\linewidth]{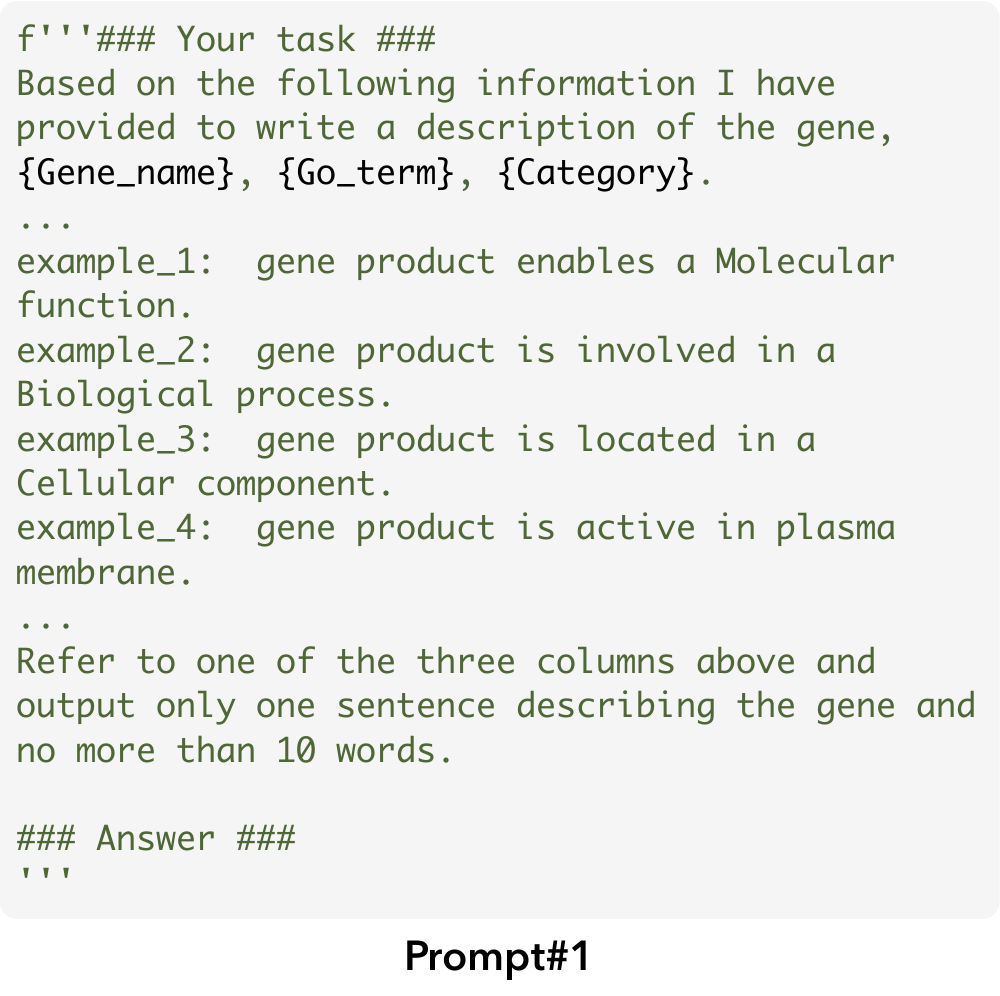}
\caption{Example template of Prompt\#1 to instruct ChatGPT for generating gene function descriptions.}
\label{prompt1}

\end{figure}
\subsection{Gene Ontology Rewrite.}
\noindent\textbf{Why We Need GO Terms Description.}
Each gene is unique due to its own functional and structural characteristics. 
GO annotations~\cite{teng2013measuring} provide gene-specific information and have proven useful for selecting Gene Reference into Function (GeneRIF) candidates. 
We aim to establish a multi-angle gene-specific description in three aspects: molecular function, biological process, and cellular component.



\noindent\textbf{Prompt-based Gene Ontology Descriptions Rewrite.}
In this study, we adopted a single-turn dialogue approach and constructed a Prompt~\cite{wang2023gpt} tailored to meet task requirements.
As shown in Figure\ref{prompt1}, this process primarily involved the following steps: 
(1) Provide task instructions. Give LLM a clear and precise task description.
(2) Provide representative examples to aid ChatGPT in fully understanding contextual semantic information and improving model performance. 
We offer four example sentences, encompassing three aspects of GO annotations.
(3) Express generation requirements to ensure standardized output. 
This paper mandates that LLM produces only one sentence per response, not exceeding 10 words, to ensure accurate expansion without additional redundant information.
Finally, we collect all generated descriptions to form a gene function description set. 

\subsection{Clustering and Streamline.}
\noindent We denote the set of filtered sentences resulting from the above steps as: $S=\left\{s_{1}, s_{2}, \cdots, s_{n}\right\}$ and the set of descriptions as: $T=\left\{t_{1}, t_{2}, \cdots, t_{m}\right\}$, where $n$ and $m$ is the size of filtered sentence set and GO term description set.

\noindent{\bf Vectorization.}
We first convert each textual data into numerical vectors. In this study, we employ BioBERT~\cite{yu2019biobert,lee2020biobert,zhu2020extracting} for vectorization, which is a domain-specific language representation model pretrained on a large biomedical corpus. 
Biomedical texts contain rich semantic information, and polysemy is common across different domains. To comprehensively capture sentence information, the encoder part of BioBERT serves as the embedding layer of the model, responsible for generating sentence vector embeddings.
BioBERT effectively captures semantic information from the text and converts it into vector representations as follows:
\begin{equation}
\begin{aligned}
V_{s_{i}}=\text{BioBERT}\left(s_{i}\right) \\
V_{t_{i}}=\text{BioBERT}\left(t_{i}\right)
\end{aligned}
\end{equation}

After vectorizing the filtered sentence set and GO terms descriptions separately, we can combine them into a global vector matrix $\mathcal{V}$:

\begin{equation}
    \begin{aligned}
        \mathcal{V} =\left\{V_{s_{1}}, V_{s_{2}}, \cdots, V_{s_{n}},V_{t_{1}}, V_{t_{2}}, \cdots, V_{t_{m}}\right\}
    \end{aligned}
\end{equation}

\smallskip
\noindent{\bf Clustering.}
We adopted K-means~\cite{steinley2006k,sinaga2020unsupervised} as the clustering method.
Each element in $\mathcal{V}$ is described by $z$ features, where $z$ is the same as the hidden side of BioBERT. 
The observation matrix of $n+m$ objects across $z$ features is structured (each row represents an object, and each column represents a feature).
The range of $k$ is set from 3 to 10. 
This choice stems from the fact that GO annotations have three aspects, thus requiring at least 3 clusters. 
Additionally, many genes have around 10 GO annotations on average. 
This approach allows for dynamic adjustment of $k$ based on the performance of clustering with different values, with the aim of achieving optimal results. 

\noindent\textit{Step-1:} The fist step involves randomly selecting $k$ points (each representing an object) from the data matrix $\mathcal{V}$ as the initial cluster centers.

\noindent\textit{Step-2:} Calculate the distance from each object and assign it to the cluster with closet center. 

\noindent\textit{Step-3:} Then we update the centroids of each cluster.

\noindent\textit{Step-4:}  Repeat the aforementioned Step-2 and Step-3 until the position of each cluster center no longer change.

\noindent\textit{Step-5:} Calculate the Calinski-Harabasz (CH) score for different values of $k$ and select the value of $k$ that yields the highest score to cluster.

\begin{figure}[t]
\centering
\includegraphics[width=0.9\linewidth]{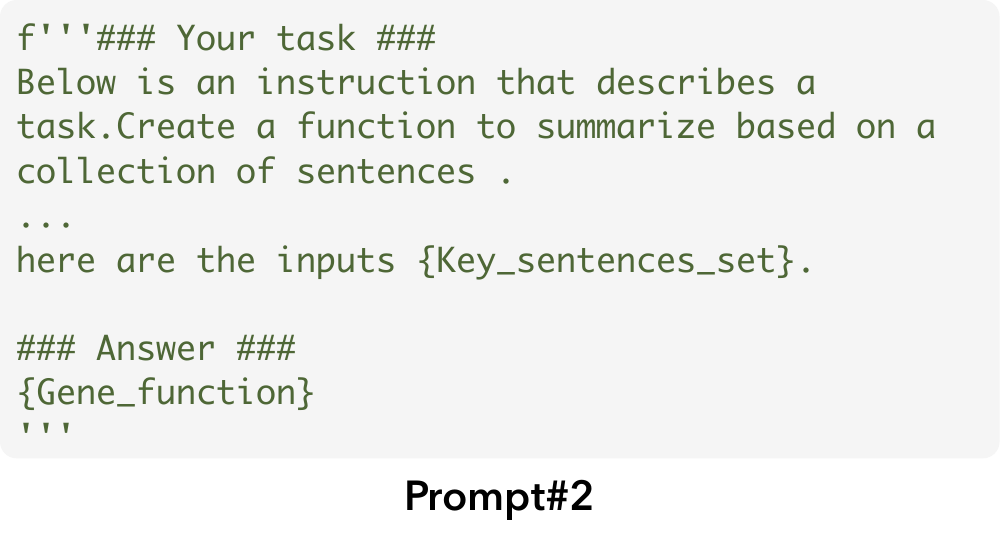}
\caption{Example of Prompt\#2 to help large language model to summary the provided knowledge.} 
\label{prompt2}
\end{figure}
\smallskip

\noindent{\bf Similarity comparison.}
At this stage, we adopt the GO term descriptions to identify the sentence with the lowest cosine distance as key sentences~\cite{rahutomo2012semantic}~\cite{xia2015learning}: 
\begin{equation}
    \begin{aligned}
        d_{i j}=\operatorname{Cosine}\left(V_{t_{i}},V_{s_{j}}\right)=\frac{V_{t_{i}}^{T} V_{s_{j}}}{\left|V_{t_{i}}\right| \times\left|V_{s_{j}}\right|},
    \end{aligned}
\end{equation}
where ${V}_{t_{i}}$ and ${V}_{s_{j}}$ are part of the same cluster. 
A higher cosine value reflects a smaller angle between the vectors, indicating better alignment and greater similarity between the n-dimensional vectors. 
Consequently, we evaluate the similarity of each sentence and description on the basis of the cosine distance and choose the sentence that shows the highest semantic similarity to the GO terms.

\subsection{Injection and Generation.}
\noindent\textbf{Method for Knowledge Injection.}
The advent of large language models has revolutionized natural language processing, offering significant advantages in text generation and adaptation. 
The introduction of LoRA (Low-Rank Adaptation)~\cite{hu2021lora} has been pivotal, reducing the number of training parameters, which reduces training time, storage, and computational demands. Combined with prefix adjustment, LoRA improves the adaptability and efficiency of the model.

We employ Gemma-7B as the base model, utilizing LoRA to fine-tune the process to generate refined summaries of genetic knowledge~\cite{pathak2023performance}. 
The process begins with the extraction of key sentences for each gene from our training dataset using a key sentence extraction module. 
Subsequently, we craft a task-specific prompt, as shown in Figure~\ref{prompt2}, which acts as a prefix during fine-tuning. This setup ensures that the model, post-LoRA fine-tuning, recognizes and efficiently executes the task of generating gene summaries.

Let the pre-training weight of Gemma-7B be $W_{0} \in R^{d \times k}$. The updates can be represented by a low-rank decomposition:
\begin{equation}
    W_{0}+\Delta W=W_{0}+W_{B} W_{A}
\end{equation}

where $W_{B} \in R^{d \times r}, W_{A} \in R^{r \times k}$, and the rank $r \ll \min (d, k)$. For a linear layer $h=W_{0} x$, the forward pass is modified to be:
\begin{equation}
    h=W_{0} x+\Delta W x=W_{0} x+W_{B} W_{A} x
\end{equation}

The matrix $W_{A}$ is initialized with random Gaussian values and $W_{B}$ is initialized to zero, setting the initial value of $\Delta W=W_{B} W_{A}$ to zero at the start of training. 
We adjusted only the attention weights for the downstream task and froze the MLP modules, applying LoRA to fine-tune all linear layers simultaneously. 


\noindent\textbf{Method for Generating Gene Summaries.}
Following the fine-tuning phase, the fine-tuned Gemma-7B model is used to generate the final gene summaries. 
Generation is driven by prompts that were used during the fine-tuning phase, ensuring consistency and relevance in the summaries produced. 
The generation process leverages the trained model's ability to synthesize information and produce output that is not only accurate, but also aligned with scientific discourse.

\section{Experimental Results}
In this section, we present detailed experimental setups and conduct comprehensive experimental analyzes and case studies to validate the efficacy of the proposed model.
\subsection{Experimental Setup.}
\smallskip
\noindent\textbf{Dataset Description.}
We utilized gene function description information from the \textit{summary} attribute of the database as a reference from NCBI sub-database Entrez Gene.
We obtained Medline PubMed IDs for all documents related to the respective genes from the PubMed data provided by Entrez Gene. 
Among numerous human-related genes, we selected 8,887 genes that had existing gene function description information for experimentation. 

\smallskip
\noindent\textbf{Evaluation Metrics}
We adopt the ROUGE-1, ROUGE-2, and ROUGE-L as metrics for evaluation as the same in~\cite{shang2014learning,lin2004rouge}. 

\smallskip
\noindent\textbf{Baseline Algorithms}
To assess the performance of our model, we selected three categories comprising six different models.
The first group of baselines is extractive-summarization approaches:
\textit{(1) Random}. This baseline randomly selects five sentences from the candidate sentences about genes and generates the description by the same LLM as \model.
\textit{(2) LTR}~\cite{shang2014learning} use three features as a basis for sentence selection: gene ontology relevance, topic relevance, and TextRank. 
The second group of baselines is General LLM:
We adopt \textit{(3) Llama2-70B}~\cite{floridi2020gpt} and \textit{(4) ChatGPT-3.5}~\cite{floridi2020gpt} with prompt and each gene's related literature as context. 
The third group of baselines is Biology-related LLM: We selected \textit{(5) BioMistral-7B-DARE}~\cite{li2024cancerllm} and \textit{(6) Llama3-OpenBioLLM-8B}~\cite{labrak2024biomistral}, large-scale biomedical models trained on meticulously curated training datasets in the field of biology and medicine. 
Following those approaches with large language models, we created prompts to enable them to automatically identify key sentences from candidate sentences about genes and generate summaries of genetic knowledge.

\smallskip
\noindent\textbf{Hyperparameter Settings and Reproducibility}
In our experiments, the significance level is set to 0.001, thus the corresponding chi-square value  is 10.83. 
Terms with a $X^{2}$ value above 10.83 would be selected as signature terms. 
We have obtained a total of 3710 terms. 
We consider all fully connected layers in gemma-7B as the target layers to be adapted and rank r=32.
Gradient accumulation step is set to 4. Gradient accumulation allows accumulating gradients over multiple batches before updating model parameters, which helps in handling larger batch sizes without consuming excessive memory. The warming steps for the learning rate are 2. The learning rate is set to 3e-4.

\smallskip
\noindent\textbf{Environmental Settings}
All experiments were conducted on the Ubuntu 18.04.6 LTS operating system, AMD EPYC 7742 CPU, and 1 NVIDIA A100 GPU, with the framework of Python 3.8.10 and PyTorch 2.0.1.
\subsection{Experimental Results}
\begin{table}[t]
\centering
\caption{Overall performance comparison for ROUGE metrics. `ROUGE-1', `ROUGE-2', and `ROUGE-L' scores are reported. The best results are highlighted in \textbf{bold}. (\textbf{Higher values indicate better performance.})}
\begin{tabular}{llll}
\toprule
Model                & ROUGE-1         & ROUGE-2         & ROUGE-L         \\ \midrule
Random               & 0.1281          & 0.0101          & 0.1083          \\
LTR                  & 0.2195          & 0.0370          & 0.1999          \\
Llama2-70B           & 0.1834          & 0.0234          & 0.1654          \\
ChatGPT3.5 *($\geq$ 200B)    & 0.2032          & 0.0299          & 0.1855          \\
BioMistral-7B-DARE   & 0.1356          & 0.0112          & 0.1236          \\
Llama3-OpenBioLLM-8B & 0.2593          & 0.0467          & 0.2416          \\\midrule
\model\ (Ours)                & \textbf{0.3874} & \textbf{0.1856} & \textbf{0.3681} \\ \bottomrule
\end{tabular}
\label{table_overall_perfCCCC}
\end{table}

\noindent\textbf{Overall Comparison}
This experiment aims to answer: \textit{Can our model effectively generate a summary of genetic knowledge?} Table~\ref{table_overall_perfCCCC} report the overall comparison results in terms of ROUGE-1-score, ROUGE-2-score, and ROUGE-L-score.
Our model significantly outperforms six baselines across three ROUGE metrics, due to leveraging the GO information of each gene for sentence selection and employing fine-tuning of a large model for sentence generation. 
Furthermore, when comparing performance ratios between different methods, ROUGE-2 shows a greater improvement compared to the other two metrics, by at least fourfold. 
Biomedical concepts typically appear in candidate sentences with multiple words (e.g.,\textit{Gene therapy},\textit{Stem cells}, \textit{Blood pressure}, \textit{Cell membrane}), and higher ROUGE-2 scores indicate our method's ability to capture this characteristic of biomedical texts. 
From this perspective, it underscores the strengths of our approach.

\smallskip
\noindent{\bf Study of the Technical Component.}
\begin{table}[t]
\centering
\caption{The influence of clustering and streamlining in \model.}
\begin{tabular}{lccc}
\toprule
Ablation &ROUGE-1  & ROUGE-2  &  ROUGE-L  \\\midrule
Random               & 0.1281          & 0.0101          & 0.1083          \\
\midrule  Filtering & 0.1650 & 0.0179 & 0.1477 \\
Filtering + Cluster & 0.1843 & 0.0190 & 0.1498 \\
Filtering + GO & 0.3857 & 0.1804 & 0.3651 \\
 Filtering + Cluster + GO (Ours) & 0.3874 & 0.1856 & 0.3681 \\
\bottomrule
\end{tabular}
\label{tableXXXXXX}
\end{table}
This experiment aims to answer: \textit{What is the impact of each technical component?} 
To answer the question, we design four different ablation variations, each adopting a different strategy to select the key sentences. 
(1) \textit{Random} refers to the same setting as the baseline \textit{Random}.
(2) \textit{Filtering} involves segmenting the literature into sentences, applying signature terms filtering, and then randomly selecting sentences to generate gene knowledge.
(3) \textit{Filtering + Cluster} refers to identifying key sentences only by clustering.
(4) \textit{Filtering + GO} refers to identifying key sentences only by the closest sentence to the GO description. 
From Table \ref{tableXXXXXX}, we can observe that our method achieved notable improvements by incorporating two technical components on top of sentence filtering. 
Besides, adding clustering resulted in a 10.4\% improvement compared with \textit{Filter} ROUGE-1. 
The underlying driver is that \textit{Filter + Clustering} enables the selection of sentences from different aspects to enrich the summarization of genetic knowledge and thus enhance the method's performance.
Furthermore, \textit{Filtering + GO} led to substantial gains: ROUGE-1 improved by 0.2207, ROUGE-2 by 0.1677, and ROUGE-L by 0.2204, compared with \textit{Filtering}. 
This phenomenon indicates that GO annotations provide comprehensive information about gene functions that can greatly improve the selection of critical sentences. 
When combining both factors (\textit{Ours}), the performance surpasses that of single-component methods. This suggests that candidate sentences rich in genetic information are likely to group with sentences expanded from GO annotations within the same category. 

\begin{figure}[!t]
\centering
\includegraphics[width=0.8\linewidth]{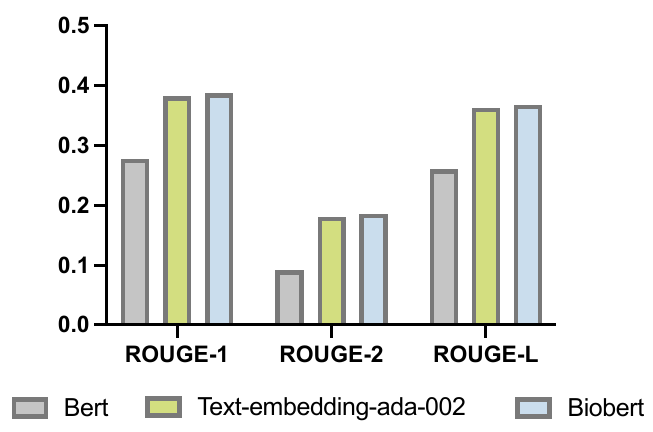}
\caption{The influence of different Vectorized models.}
\label{table1}
\vspace{-0.5cm}
\end{figure}
\smallskip

\noindent{\bf Study of the Vectorized Model.}
This experiment aims to answer the following question: \textit{What effect do different vectorization models have on the results?}
We conducted experiments on four vectorization models, including Bert~\cite{devlin2018bert}, BioBERT, and GPT-3.5 (Text-embedding-ada-002), respectively.
Through exploration with different vector models, Figure~\ref{table1} shows that GPT-3.5, which has a large parameter size, outperforms Bert. 
This suggests that GPT-3.5 has superior abilities in representing the semantic relationships between words, enhancing the generated results' quality. 
BioBERT, trained on PubMed abstracts and PMC full-text articles, is closer to our task and benefits significantly from a training dataset five times larger than Bert's, resulting in notable performance improvements across all three metrics compared to Bert.
Furthermore, we observed that even though GPT-3.5 has a larger parameter size, its performance is slightly lower than that of BioBERT, which is specifically tuned on a biomedical dataset. 
This result indicates that domain-specific language models, which are fine-tuned on relevant biomedical literature, may offer more precise and effective semantic representations for tasks closely aligned with their training data.

\noindent{\bf Study of the Impact of LLMs Selection.}
\begin{figure}[!t]
\centering
\includegraphics[width=0.8\linewidth]{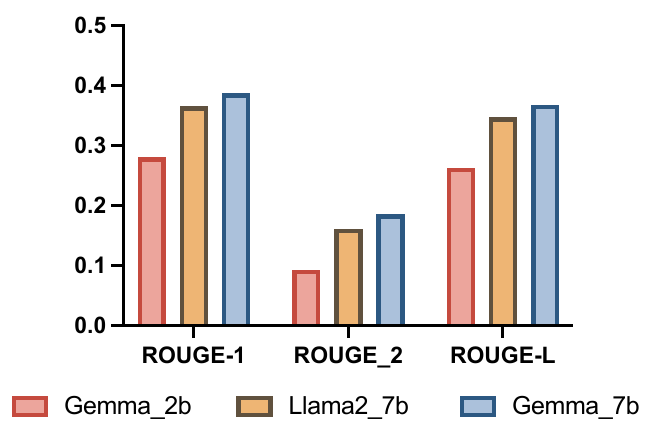}
\caption{The influence of different generative model.}
\label{table2}
\vspace{-0.5cm}
\end{figure}
This experiment aims to answer: \textit{What are the differences in performance among different open-source LLM?} 
Performance was compared in all experiments using Gemma-2B, Gemma-7B, and Llama2-7B.
From Figure~\ref{table2}, we can observe that models with larger parameters, Gemma-7B and Llama2-7B, outperformed the smaller Gemma-2B. This is because larger parameter sizes enable models to learn and adapt to data, improving performance. 
Additionally, Gemma-7B performed better than Llama2-7B, which is consistent with Google's technical report on Gemma, highlighting its superior performance in mathematics, reasoning, and code compared to similarly sized Llama2 models. 
In conclusion, a better large language model can effectively assist in generating summaries of genetic knowledge using key sentences.

\begin{figure}[t]
\centering
\includegraphics[width=1\linewidth]{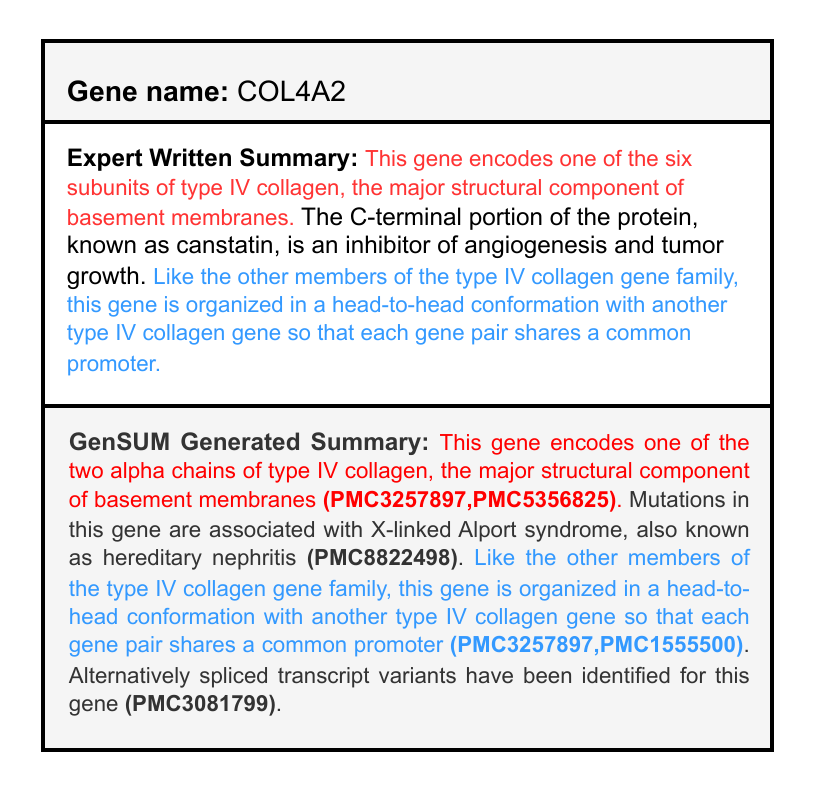}
\caption{Comparison of Generated Gene Summary for COL4A2.}
\label{table_col}
\end{figure}

\begin{figure}[t]
\centering
\includegraphics[width=1\linewidth]{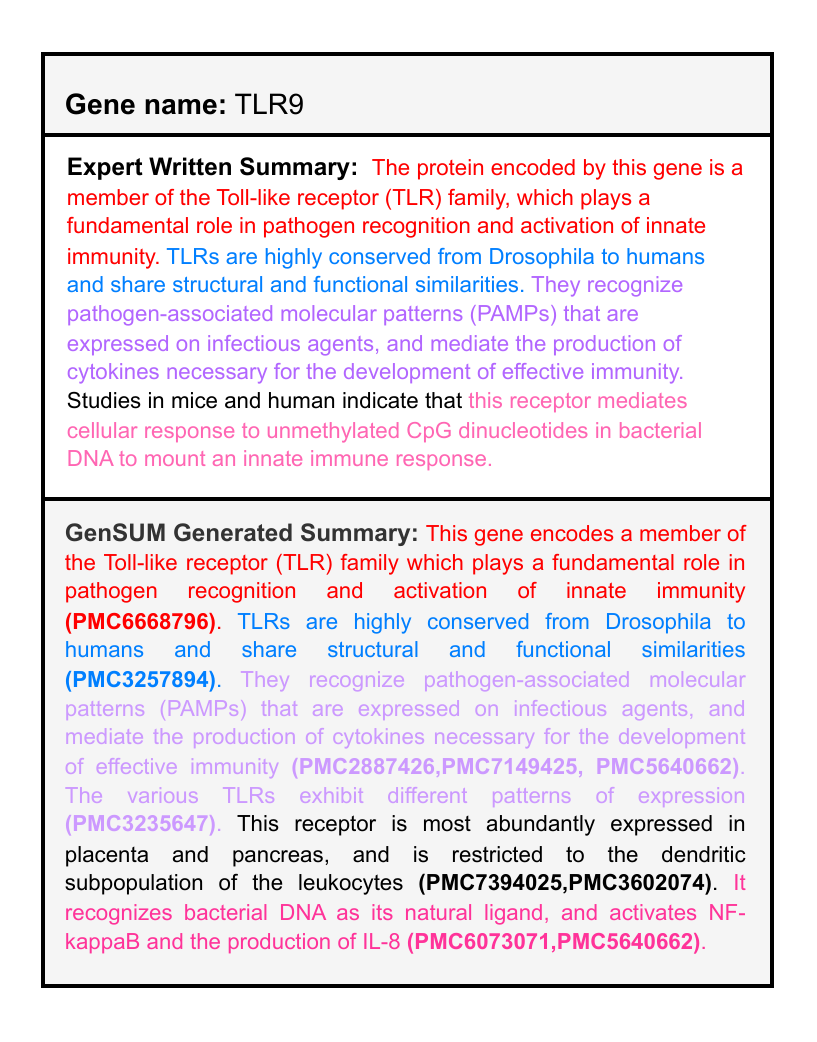}
\caption{Comparison of Generated Gene Summary for TLR9.}
\label{table_TLR}
\end{figure}

\noindent{\bf Case Studies on Generated Gene Summary.}
We further validated the generated gene summary by comparing the expert-based gene summary and the summary generated by \model.
As shown in Figure\ref{table_col} and Figure\ref{table_TLR}, we illustrate the generated gene summary for COL4A2 and TLR9. 
We also highlighted the sentences from each sample in the same color if they are semantically similar. 
Firstly, we can observe that our generated summary is highly similar to the standard summaries. 
In addition, each generated sentence can also be traced back to specific articles (marked by PubMed ID), addressing the issue of LLM hallucinations and demonstrating complete traceability. 
Finally, regardless of the length of the standard summaries, we can generate summaries of similar lengths, eliminating the influence of length on generated summaries and highlighting the advantages of fine-tuning LLMs. These examples illustrate the strengths of our two-stage method.

\section{Related Work}
The field of genetic knowledge has evolved through various methodologies, mainly classified as:
Extractive summarization techniques focus on selecting key sentences~\cite{ling2007generating} or fragments directly from the text without altering the original wording by semantic similarity~\cite{jin2009towards} reinforcement learning~\cite{chen2018fast}, information-theoretic~\cite{west2019bottlesum} and learning-to-rank~\cite{shang2014learning}. 
Generative summarization~\cite{gehrmann2018bottom} involves rewriting or generating new sentences that encapsulate the meaning of the original text, often providing more coherent and concise summaries. 
Those approaches underscore the integration of multiple scoring features to refine the selection process in extractive summarization, which is limited by the generative ability of the underlying language model. 
In the current era of large language models~\cite{ye2023needed,yuan2024continued,wang2024biorag,li2024scinterpreter,cai2023resolving}, there is a significant shift towards more sophisticated summarization techniques~\cite{floridi2020gpt,liu2021makes}, particularly in handling complex genetic data~\cite{zhang2024enhanced}. 
We propose leveraging a Language Model (LLM) to expand Gene Ontology (GO) annotations into complete sentences. This method utilizes similarity measures to pinpoint key sentences, ensuring that the summaries are not only relevant but also concise and coherent.



\section{Conclusion and Remarks}
In this paper, we address the challenge of efficiently summarizing the extensive and rapidly expanding literature on gene functions, characteristics, and expressions. 
To overcome these challenges, we introduce a two-stage automated gene summary extractor, \model, utilizing a large language model (LLM). This system initially refines the literature retrieval process to reduce redundancy and subsequently employs fine-tuning to enhance the summarization output. 
These advancements signify a substantial step forward in automating and improving the accessibility of gene-related knowledge.




\section{Acknowledgements}
This work is partially supported by the Postdoctoral Fellowship Program of CPSF (No.GZC20232736), the China Postdoctoral Science Foundation Funded Project (No.2023M743565), the Strategic Priority Research Program of the Chinese Academy of Sciences
XDB38030300.


\balance
\bibliographystyle{IEEEtran}   
\bibliography{reference}
\end{document}